\documentclass{icrc2009}

\usepackage{graphicx}   
\usepackage[caption=false]{caption}    
\usepackage[font=footnotesize]{subfig} 
\usepackage{fixltx2e}
\usepackage{url}

\newcommand{\shorttitle}[1]%
{\markboth{Proceedings of the 31\MakeLowercase{$^{st}$} ICRC, {\L}\'{o}d\'{z} 2009}{#1} }
\newcommand{\etal}{\MakeLowercase{\textit{et al. }}} 
\def\aap{{\em A\&A}}

\begin{document}
\title{Search for gamma-ray emission from solar system bodies with \emph{Fermi}-LAT}

\author{\IEEEauthorblockN{Nicola Giglietto\IEEEauthorrefmark{1} on behalf of the { Fermi}-LAT Collaboration
			  }
                            \\
\IEEEauthorblockA{\IEEEauthorrefmark{1}Dipartimento Interateneo di Fisica di Bari and INFN Bari
     }
}
\shorttitle{N.Giglietto \etal {\it Fermi} planets gamma-ray observations}
\maketitle

\begin{abstract}
 Fermi LAT is performing an all-sky gamma-ray survey from 30 MeV to $>$300 GeV with unprecedented sensitivity and angular resolution.
Fermi has detected high-energy gamma rays from the Moon and Sun since the first weeks of data taking. This emission is produced by interactions of cosmic rays with these objects. Similarly, some gamma ray emission can be produced by interactions with asteroids and planets. We have searched this emission looking major planets during the first 6 months of data taking. We present here the status of the search.
  \end{abstract}

 
\section{Introduction}
 \emph{Fermi} was successfully launched from Cape Canaveral on the 11th of June 2008.
 It is currently in an almost  circular orbit around the Earth at
 an altitude of 565~km having an inclination of 25.6$^\circ$ and an orbital period of 96 minutes. After an initial period of engineering data taking and on-orbit calibration \cite{LATcalib}, the observatory was  put into a sky-survey mode.
The observatory has two instruments onboard,
the Large Area Telescope (LAT)\cite{LATpap},  a pair-conversion gamma-ray detector and tracker   
  and a Gamma Ray Burst Monitor (GBM), dedicated to the detection of gamma-ray bursts. The instruments  on \emph{Fermi} provide coverage over  the energy range measurements from few kev to several hundreds of GeV. 

Here we report the results of a search for a gamma-ray diffuse component emissions due to major planets,
giovian trojans, neptunian troyans, asteroids and debris  of solar system. 
The cosmic-ray interactions with planets, asteroids, and objects in the Oort Cloud should produce a gamma-ray component due to the $\pi_0$ decays coming from the hadronic interactions by cosmic-rays hitting the surface of these bodies and any massive object of the solar system. Some recent works\cite{Igor3, IgorHEAD} explore the possibility that this emission should be taken into account, when dealing about gamma-ray diffuse emission, and try to compute gamma-ray fluxes for a lot of interesting class of bodies, including comets.
In particular the flux computations for the albedo gamma-ray emission from the Moon\cite{Igor, Igor2} is taken as  a template to extrapolate the emission from  solid bodies, potentially able to emit albedo gamma-rays. 
Other observations\cite{sheppard} have 
reported visible and infrared observations and position calculations about the giovian trojan establishing the diameter distribution and albedo measurements of these objects.
Moreover in  other works\cite{Moskalenko:2009tv} these calculations are now including cosmic ray interactions with debris in the Oort Cloud and concluding that probably a fraction of the previously known as extragalactic$\gamma$-ray background could be due to these objects. 

However while the emission from Moon\cite{EGRET} and Sun\cite{EGRET,IC-Orlando,OrlandoStrong} was previously detected  by EGRET, and confirmed by \emph{Fermi} \cite{icrc2009-giglietto,icrc2009-orlando}, 
 the gamma-ray flux from these interactions was never detected and should present an intensity level too low to be detected as pointlike sources. Therefore this emission should give a contribution to the extragalactic  isotropic gamma-ray background. It is straightforward that in order to
correctly evaluate the true extragalactic and galactic components of the diffuse emission, it's important to quote or to estimate upper limits of this local solar system component of gamma-ray emissions.
The first 6 months of data sample are here used to look directly the most promising regions for this emission and the first preliminary results are here presented. 

\section{Data selection}

 The data sample used includes the scientific data collected since 4 August 2008  to the end of February 2009 (7months). We have
applied a zenith cut of 105$^{\circ}$ to eliminate photons from the Earth's limb. We use for this analysis the "Diffuse" class
\cite{LATpap}, corresponding to the events with the highest probability candidates as photons. Science Tools version used is v9r11 and IRFs(Instrumental Response Functions) version P6\_V3.

 \begin{figure*}[th]
   \centering
  \includegraphics[height=4in,width=5in]{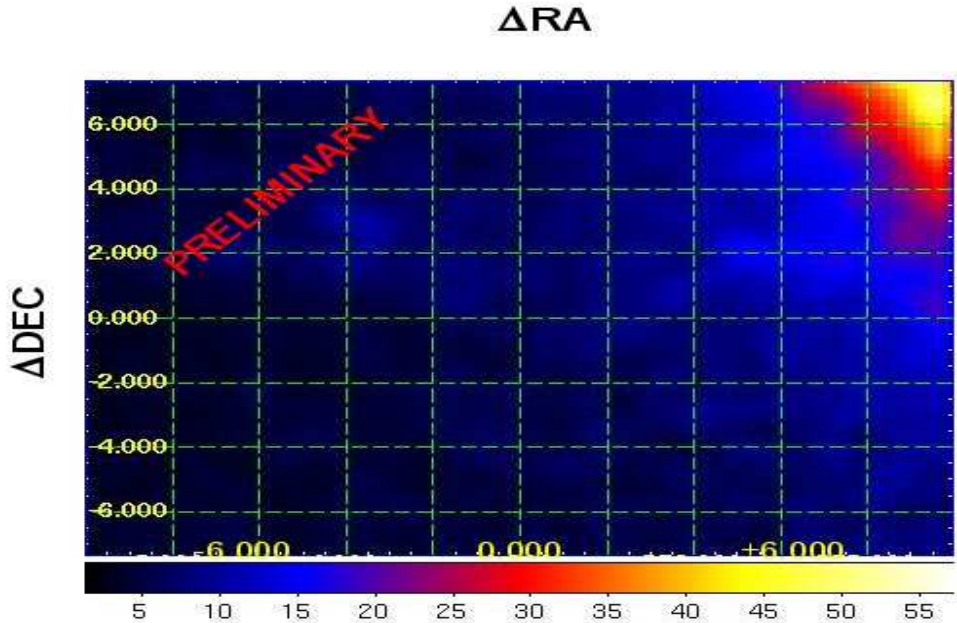}
  \caption{Count map in the Jupiter region. Coordinates are celestial offset in degrees respect to the Jupiter position. The bin width used is 0.2$^\circ$ and the image is obtained 
using a colour scale linearly proportional to the counts; a  gaussian smoothing  using a kernel radius of 3, has
 then applied to the image. On the top right corner of the image is partially visible the galactic plane emission.}
\label{jupiter-map}
  
\end{figure*}
The best way to observe the  $\gamma$-ray emission coming from the direction of major planets is to look the count maps in that regions. In most situations the planet displacement in the sky may be considered negligible and the standard \emph{Fermi} Science Tools can be used.
 For any other case we apply the same tools developed for solar system bodies.
 We have then selected events in celestial coordinates relative to  the position of major planets. Moreover we have tried the exploration of the ecliptic regions. For any selected object we have examined the count map to look for any evident excess of counts.

\section{Results}
For this analysis we report only qualitative results about the $\gamma$-ray emission from large planets. Fig.~\ref{jupiter-map} shows the count map of the events in celestial coordinates offsets relative to Jupiter position.
We are starting to  carefully quote the fluxes or the upper limits of the fluxes from these objects. However for this analysis, taking into account that the 
integrate sensitivity in the first 6 months should not be sufficient to have a clear evidence of emission from weakest sources and taking into account that
Saturn and Jupiter has spent most of the time in regions close to the galactic plane, no clear emission centered on these planets is visible or within few degrees from these sources. 
We have moreover reviewed the regions close to Uran and Neptune too, and no signal was seen during the first six months. 
%
%
%
\section{Conclusions}
We have searched for the gamma-ray albedo emission from Jupiter.   Using the first 6-months data we have no evidence of gamma-ray emission from this object. A detailed analysis
  search for emission from other bodies of the solar system objects, in particular from Saturn region, is in progress. 
However the sensitivity of \emph{Fermi} to detect weak sources bring to conclude that increasing the statistics and the exposures, we should have the possibility to reveal at least the unresolved diffuse emission connected to these bodies.

\section{Acknowledgements}

The $Fermi$ LAT Collaboration acknowledges support from a number of agencies and institutes for both development and the operation of the LAT as well as scientific data analysis. These include NASA and DOE in the United States, CEA/Irfu and IN2P3/CNRS in France, ASI and INFN in Italy, MEXT, KEK, and JAXA in Japan, and the K.~A.~Wallenberg Foundation, the Swedish Research Council and the National Space Board in Sweden. Additional support from INAF in Italy for science analysis during the operations phase is also gratefully acknowledged.

\end{document}